\documentclass[superscriptaddress, reprint, amsmath, amssymb, aps,
prl]{revtex4-1}

\usepackage{graphicx}%
\usepackage{xspace}%
\usepackage{url}%

\usepackage{color}

\newcommand{\musrfit}{\texttt{musrfit}\xspace}
\newcommand{\musrview}{\texttt{musrview}\xspace}
\newcommand{\msrtodata}{\texttt{msr2data}\xspace}
\newcommand{\musreditgui}{\texttt{musredit}/\texttt{musrgui}\xspace}
\newcommand{\minuit}{\texttt{Minuit2}\xspace}
\newcommand{\musr}{$\mu$SR\xspace}
\newcommand{\lem}{LE-$\mu$SR\xspace}
\newcommand{\mup}{$\mu^+$\xspace}
\newcommand{\ROOT}{\texttt{ROOT}\xspace}
\newcommand{\ascii}{\texttt{ASCII}\xspace}
\newcommand{\cpp}{\texttt{C++}\xspace}

\begin{document}

\title{\musrfit: a free platform-independent framework for \musr data analysis}

\author{A.~Suter}
\email{andreas.suter@psi.ch}
\affiliation{Labor f\"ur Myonspinspektroskopie, Paul Scherrer Institut, 5232
Villigen PSI, Switzerland}

\author{B.~M.~Wojek}
\affiliation{Physik-Institut, Universit\"at Z\"urich, 8057 Z\"urich,
Switzerland}
\affiliation{Labor f\"ur Myonspinspektroskopie, Paul Scherrer Institut, 5232
Villigen PSI, Switzerland}

\begin{abstract}
A free data-analysis framework for \musr has been developed. \musrfit is fully
written in \cpp, is running under GNU/Linux, Mac OS X, as well as Microsoft
Windows, and is distributed under the terms of the GNU GPL. It is based on the
CERN \ROOT framework and is utilizing the \minuit optimization routines for
fitting. It consists of a set of programmes allowing the user to analyze and
visualize the data. The fitting process is controlled by an \ascii-input file
with an extended syntax. A dedicated text editor is helping the user to create
and handle these files in an efficient way, execute the fitting, show the data,
get online help, and so on. A versatile tool for the generation of new input
files and the extraction of fit parameters is provided as well.
\musrfit facilitates a plugin mechanism allowing to invoke user-defined
functions. Hence, the functionality of the framework can be extended with a
minimal amount of overhead for the user. Currently, \musrfit can read the
following facility raw-data files: \texttt{PSI-BIN}, \texttt{MDU} (PSI),
\texttt{ROOT} (LEM/PSI), \texttt{WKM} (outdated \ascii format), \texttt{MUD}
(TRIUMF), \texttt{NeXus} (ISIS).
\end{abstract}

\pacs{76.75.+i \sep 07.05.Kf \sep 07.05.Rm}
\keywords{$\mu$SR data analysis}

\maketitle

\section{Introduction}

Currently, various data-analysis tools for analyzing muon-spin-rotation (\musr)
data are available, however, the situation is unsatisfactory for different
reasons. Most of these programmes are limited to a single platform,
\emph{e.\,g.} Microsoft Windows, others are not maintained anymore, and some of
them are not free software. Another problem arising from this situation is that
more elaborate modelling is almost impossible from within the current frameworks
and hence, the users are forced to write their own code. This was especially
true for low-energy \musr (\lem) where often the \mup stopping distribution has
to be taken into account in the analysis. Therefore, we started to develop a
free data-analysis framework for \musr, called \musrfit, which should overcome
the problems described.

From the points raised, the design criteria were: (i) \musrfit has to be free
software according to the GNU licenses \cite{GPL}, and hence available to
everyone. (ii) It should be transparent and user-friendly, \emph{e.\,g.} a clear
and complete online documentation should be available. (iii) Extensions to the
basic framework should be possible on the user level. (iv) \musrfit should be
able to read all currently used \musr-data-file formats directly. (v)
Data-visualization and fit-parameter tools should be available. (vi) The
maintainability should be warranted.

In order to fulfil all these requirements we decided to build up on the \ROOT
framework \cite{ROOT} developed and maintained at CERN and heavily used in
particle physics as well as other fields of physics and engineering. The \ROOT
framework is a collection of \cpp libraries together with a \cpp macro
interpreter. It provides graphical-user-interface tools and contains the
\texttt{Minuit}/\minuit optimization routines \cite{Minuit}. For us it is a
``natural'' choice since the \ROOT framework is already part of the
\musr-data-acquisition systems at PSI. \musrfit consists of a collection of \cpp
classes \cite{musrfit-tech} which can be used either directly using the \ROOT
macro capabilities, or---probably simpler for most of the users---can be
accessed via some user-friendly programmes, provided within the \musrfit suite
described in the next section.

\section{The \musrfit suite}

The analysis of \musr data using the \musrfit suite is controlled by text files
with the extension \texttt{msr} (``\texttt{msr} files''). These human-readable
files contain all information needed to fit a model function to the \musr data:
the fit parameters, the definition of the model, some details on the relevant
\musr data files, and the fitting routines to be used. Moreover, information
used for the graphical presentation of the data and fits, such as plot ranges
and parameters for Fourier transforms are stored in the \texttt{msr} files as
well. These \texttt{msr} files are also used as a protocol of the fit results. A
detailed explanation of the structure and syntax of the \texttt{msr} files can
be found in Ref.~\cite{musrfit-doc}. In the following, only a basic overview of
the different programmes shall be given.

\subsection{\musrfit---fitting a model}
\begin{figure}[ht!]
\centering
\includegraphics[width=.45\textwidth]{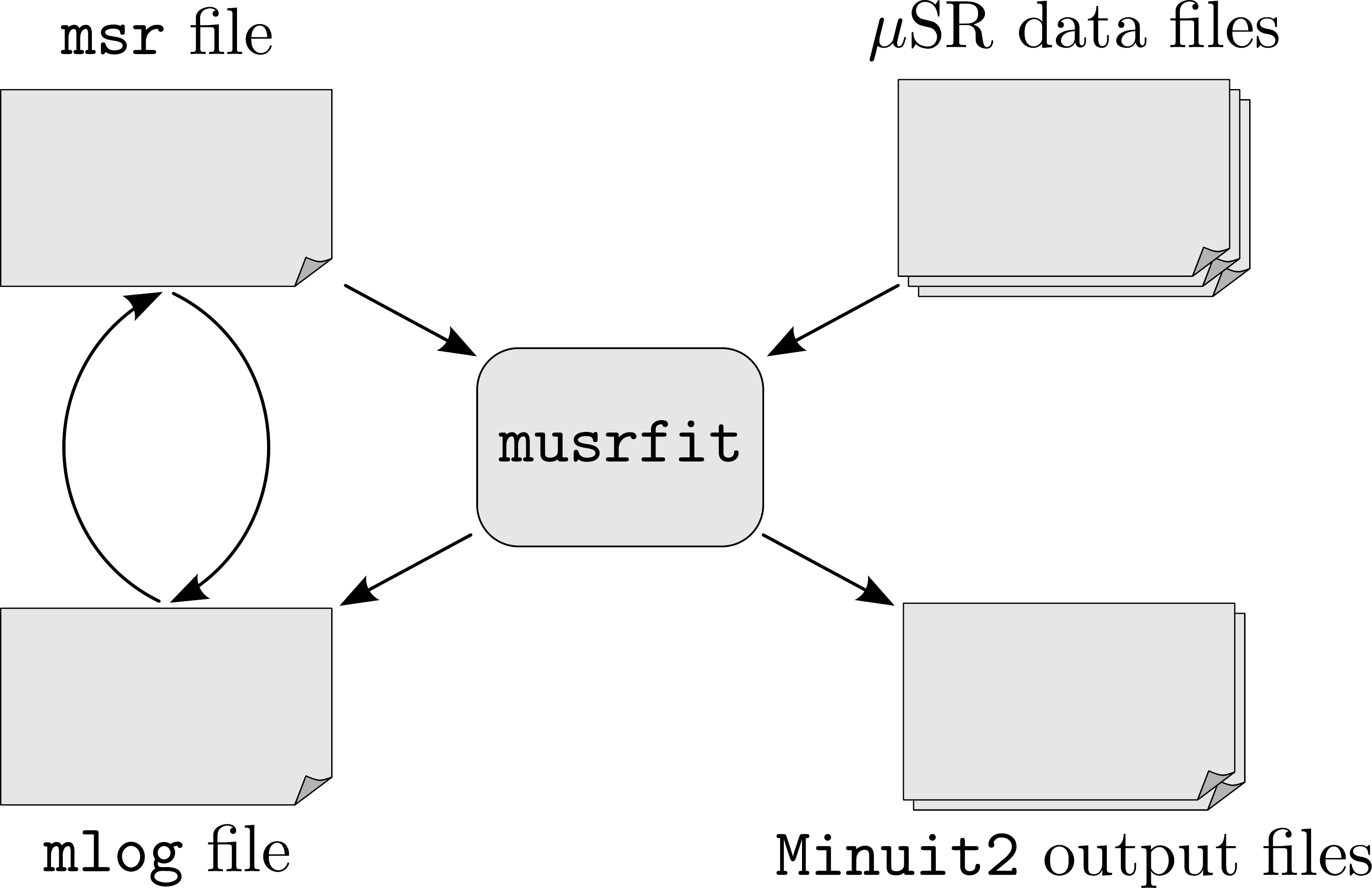}
\caption{General file flow during a fit using \musrfit.}
\label{fig:musrfit}
\end{figure}

After \musrfit is called to fit a model it analyzes the respective \texttt{msr}
file and reads in all specified data files. Successively, the fit is performed
and the resulting parameters are written to a \texttt{mlog} file which also
complies with the \texttt{msr} file structure. Additionally, the covariance
matrix and the correlation coefficients of the free fit parameters as determined
by \minuit are saved as \ascii and binary \ROOT files. In a final step, the
\texttt{msr} and \texttt{mlog} files are swapped so that the \texttt{msr} file
contains the updated parameter values while the \texttt{mlog} file holds a copy
of the parameter set used as input to \musrfit. This procedure is summarized in
Fig.~\ref{fig:musrfit}. Currently, \musrfit supports $\chi^2$ minimization and
log-likelihood maximization.

\subsection{\musrview---graphical presentation}
\begin{figure}[ht!]
\begin{center}
\includegraphics[width=.495\textwidth]{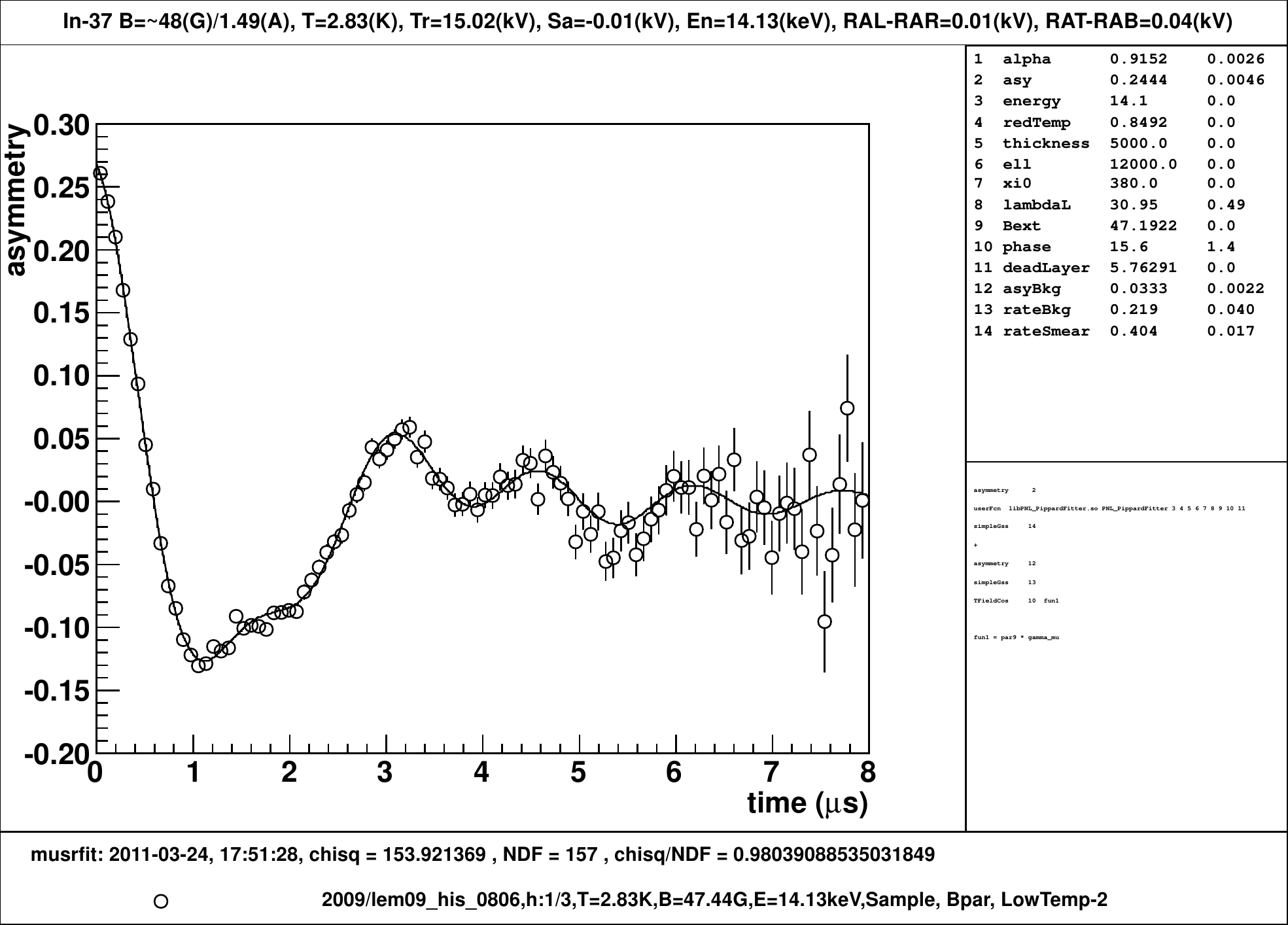}
\includegraphics[width=.495\textwidth]{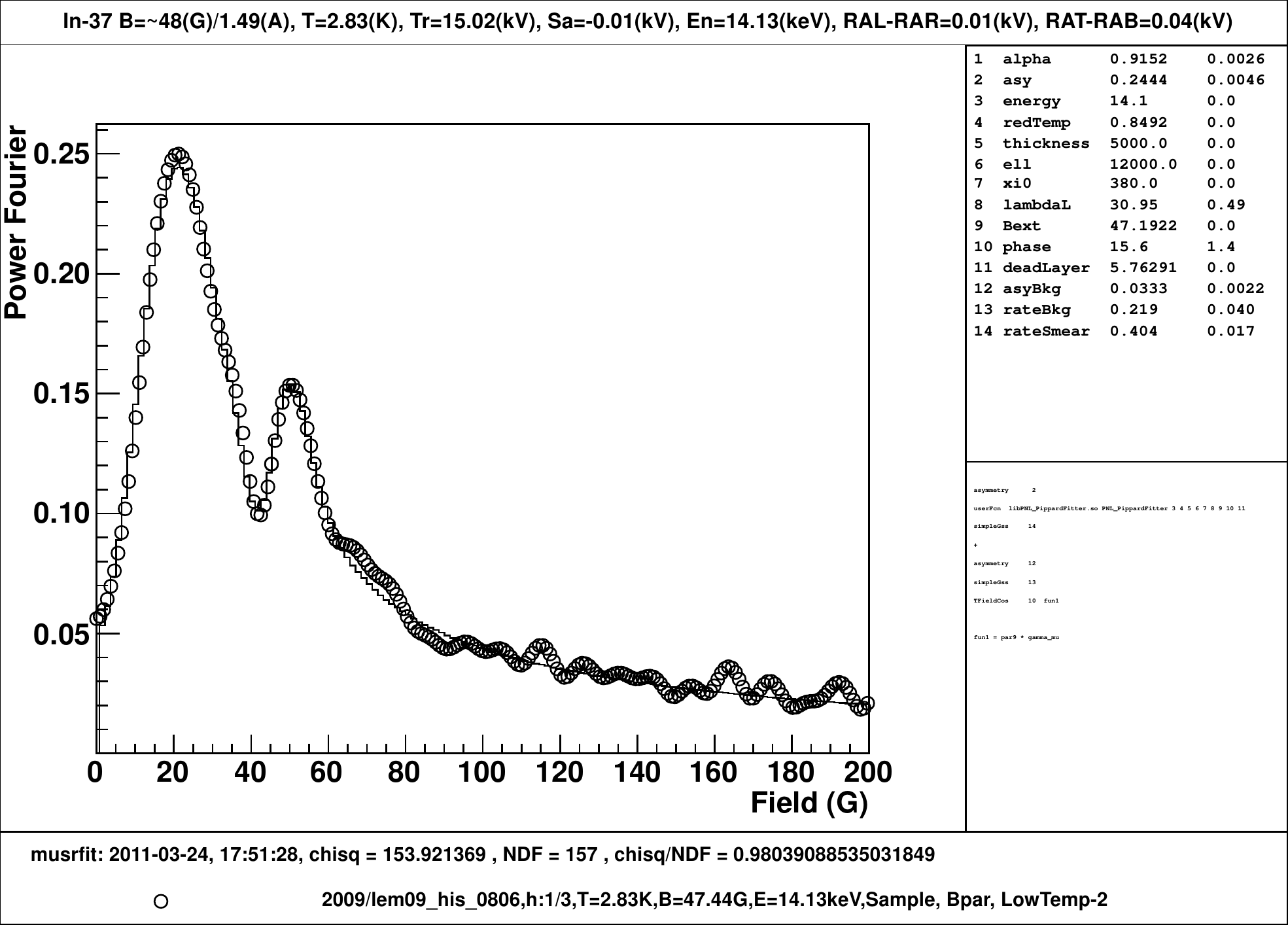}
\end{center}
\caption{\emph{Left panel:} Time spectrum of selected \musr data plotted using
\musrview. \emph{Right panel:} Corresponding magnetic-field distribution. In
both cases, the solid lines represent the fitted model. Both graphs contain
information on the fit parameters (upper right box), the model function (lower
right box), the fit statistics as well as the plotted data (bottom box).}
\label{fig:musrview}
\end{figure}

The analyzed data and the model functions can be visualized using the programme
\musrview. Like the actions of \musrfit also the initial plotting frame of
\musrview is controlled by the \texttt{msr} file. For example, it can be
specified which data should be drawn in which range; if more than one set of
data should be drawn it can be chosen if they should appear in a single canvas
or in separate ones, and so on. Since the plotting routines are based on \ROOT
as well, the initially drawn graphs can be easily modified using conventional
\ROOT editing features---\emph{e.\,g.} labels or arrows could be added, colours
could be changed, and so on. \musrview also offers the possibility to calculate
and show the Fourier transforms of \musr time spectra. As an example,
Fig.~\ref{fig:musrview} depicts the time spectrum and the corresponding field
distribution of a selected \musr measurement plotted by \musrview. Despite not
being shown in Fig.~\ref{fig:musrview}, also the difference between the analyzed
data and a given model can be plotted. Furthermore, a set of keyboard shortcuts
has been implemented to make the navigation more easy, \emph{e.\,g.} pressing
`\texttt{f}' toggles between the data presentation in the time and frequency
domains, `\texttt{d}' changes the view to the difference plot. Finally, there
also exists the possibility to save the shown data and model curves in an \ascii
file which facilitates the further use of these data in the user's favourite
programme.

\subsection{\msrtodata---advanced \texttt{msr}-file handling}
The \musrfit suite also hosts a tool called \msrtodata. Its main purpose is to
process multiple \texttt{msr} files with the same parameters and to summarize
the fit parameters contained in the \texttt{msr} files either in a TRIUMF
\texttt{DB} file~\cite{triumfDB} or a column \texttt{ASCII} file. Moreover,
\msrtodata can be used to generate from a template new \texttt{msr} files and
even a ``global'' \texttt{msr} file for various runs sharing a subset of common
parameters; for details on the ``global'' \texttt{msr}-file handling refer to
Ref.~\cite{musrfit-doc}.
\begin{figure}[ht]
\centering
\includegraphics[width=0.8\linewidth]{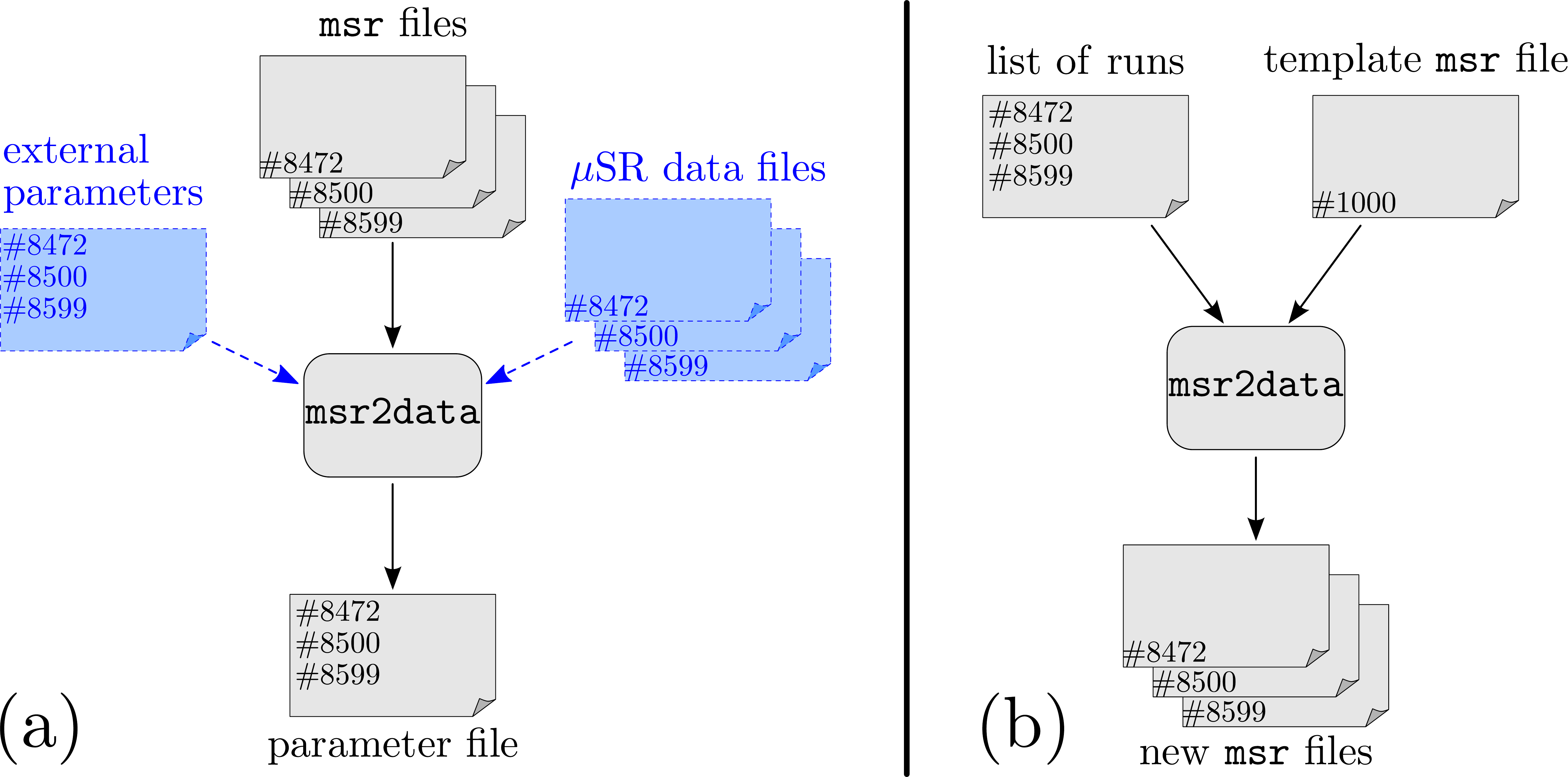}
\caption{(a) \msrtodata used for extracting fit parameters from a list of
\texttt{msr} files and summarizing them in a parameter file. The files sketched
in blue with broken frames can be specified optionally. (b) \msrtodata used for
generating new \texttt{msr} files from a template.}
\label{fig:msr2data}
\end{figure}\\
Figure~\ref{fig:msr2data}(a) shows schematically the parameter extraction from
different \texttt{msr} files. \msrtodata is provided with a list of runs to be
processed; optionally, external parameters which should be included in the
resulting parameter file can be specified for each of the runs. Also optionally,
parameters potentially stored in the \musr data files (temperature, applied
magnetic field, and so on) can be asked to be included. \msrtodata then reads
the \texttt{msr} files for all given runs and adds all the parameter information
to a parameter file. Figure~\ref{fig:msr2data}(b) illustrates the
\texttt{msr}-file generation using a template---essentially, new \texttt{msr}
files are created by substituting the run number in the template.\\
It is possible as well to combine the above described actions: a template can be
used to create new \texttt{msr} files for a list of runs, these files in turn
are processed by \musrfit, and finally the parameters of each of the files are
summarized automatically in a parameter file. A detailed description of all
possible options can be found in Ref.~\cite{musrfit-doc}.

\subsection{\musreditgui---editing \texttt{msr} files}
\begin{figure}[bt]
\centering
\includegraphics[width=0.8\linewidth]{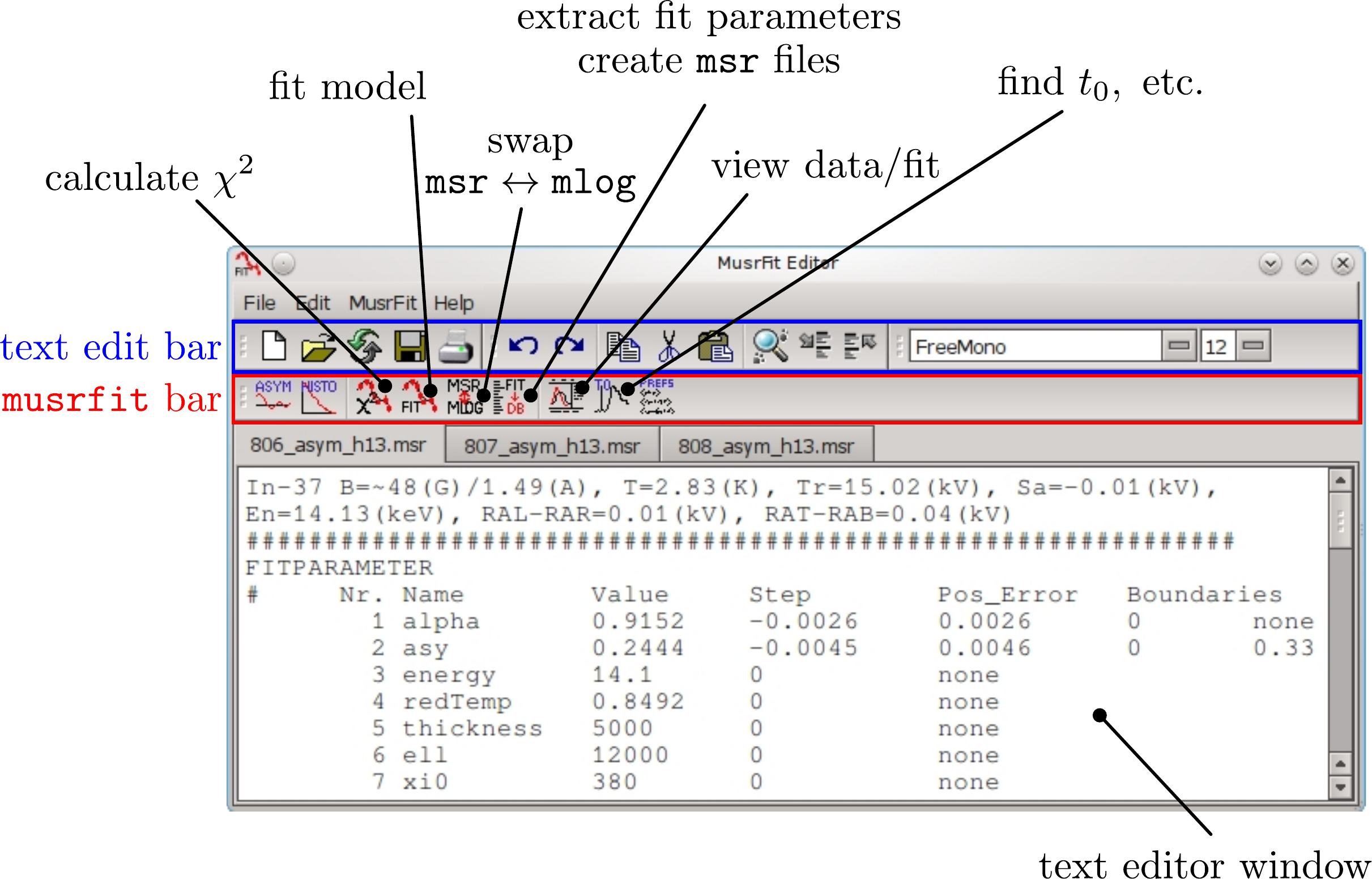}
\caption{Screenshot of a \texttt{musredit} window. The most important \musrfit
features are highlighted.}
\label{fig:musredit}
\end{figure}

Even though the \texttt{msr} files can naturally be edited with any text editor
and the various programmes of the \musrfit suite can be called from the command
line, with \musreditgui dedicated text editors which also serve as frontends for
the \musrfit framework are provided. These are specifically intended to help the
user handle \texttt{msr} files. Principally, \texttt{musredit} and
\texttt{musrgui} have the same capabilities, however, they are based on
different versions of \texttt{Qt}~\cite{Qt}: \texttt{musrgui}---\texttt{Qt\,3},
\texttt{musredit}---\texttt{Qt\,4.6} or newer. Both programmes feature basic
editor functions as well as interfaces to \musrfit and are documented in
Ref.~\cite{musrfit-doc}. A screenshot of \texttt{musredit} indicating the most
important \musrfit features accessible through the editor is shown in
Fig.~\ref{fig:musredit}.

\subsection{User-defined functions}
Additionally to providing a set of pre-defined muon-spin-polarization functions,
the \musrfit suite facilitates a plug-in mechanism allowing it to invoke
user-defined functions implemented in \cpp classes~\cite{musrfit-doc}. Hence,
the functionality of the programmes can be extended with a minimal amount of
overhead for the user. For example, for the analysis of \lem data it is
sometimes desirable to take into account the muon stopping distribution in the
calculation of the depolarization function---the plug-in mechanism offers a
possibility to do so. A couple of such add-on modules are already available
together with \musrfit, \emph{e.\,g.} for modelling data obtained from
superconductors which exhibit local~\cite{Kiefl-PRB-2010} or
nonlocal~\cite{Suter-PRB-2005} Meissner screening of an applied magnetic field
below their surfaces. It should be noted as well, that the so-defined plug-in
classes are of course not limited to the use within \musrfit; for instance, one
could think about reusing them in other programmes or \ROOT macros.

\subsection{any2many---a ``universal'' \musr-data-file converter}
The \musrfit suite reads the currently available \musr data files without any
conversion necessary. However, users might favour their own analysis software
but have difficulties with all the different \musr facility data-file formats.
Hence, the small helper programme \texttt{any2many} is included in the \musrfit
suite which allows virtually all possible conversions from one format into
another, including \ascii output.

\section*{Acknowledgements}
We are indebted to Z.~Salman, T.~Prokscha, H.~Luetkens, and R.~Scheuermann for
helpful discussions, suggestions, and bug reporting. Thanks to D.~Arseneau for
releasing \texttt{MUD} under the GNU LGPL~\cite{GPL}, and S.~Cottrell for the
discussion and information provided on \texttt{NeXus}.

\end{document}